\documentclass{aa}
\usepackage{graphicx}
\usepackage{txfonts}
\usepackage{xspace}
\usepackage{xcolor}
\usepackage{url}
\usepackage{hyperref}
\usepackage{orcidlink}
\usepackage[switch]{linenoaa} 
\newcommand {\flux}{erg~s$^{-1}$~cm$^{-2}$\xspace}
\newcommand {\lum}{erg\,s$^{-1}$\xspace}

\newcommand{\jsixt}{\mbox{IGR~J16418$-$4532}\xspace}
\newcommand{\jelev}{\mbox{IGR~J11215$-$5952}\xspace}
\newcommand{\jsevent}{\mbox{IGR~J17544$-$2619}\xspace}

\begin{document}

\title{Low-state X-ray spectra of magnetized neutron stars: Probing accretion regimes in SFXTs}

\author{
Hua~Xiao \inst{\ref{in:UTU}}\orcidlink{0009-0004-1288-4912}
\and Sergey~S.~Tsygankov \inst{\ref{in:UTU}}\orcidlink{0000-0002-9679-0793}
\and Alexander~Salganik \inst{\ref{in:UTU}}\orcidlink{0000-0003-2609-8838}
\and Alexander~A.~Mushtukov \inst{\ref{in:UCL},\ref{in:OXF}}\orcidlink{0000-0003-2306-419X}
\and Juri~Poutanen \inst{\ref{in:UTU}}\orcidlink{0000-0002-0983-0049}
          }
          
\institute{Department of Physics and Astronomy, FI-20014 University of Turku,  Finland \label{in:UTU} \\ \email{hua.h.xiao@utu.fi}
\and 
Mullard Space Science Laboratory, University College London, Holmbury St. Mary, Surrey RH5 6NT, UK \label{in:UCL}
\and
Astrophysics, Department of Physics, University of Oxford, Denys Wilkinson Building, Keble Road, Oxford OX1 3RH, UK
\label{in:OXF}
}
          
\titlerunning{Low-state spectra of SFXTs}
\authorrunning{Xiao et al.}


\abstract{
Supergiant fast X-ray transients (SFXTs) are a subclass of high-mass X-ray binaries characterized by prolonged quiescent states, making them ideal sources for studying accretion onto highly magnetized neutron stars (NSs) at low rates. In this work, we present a broadband spectral analysis of three SFXTs: \jsixt, \jsevent, and \jelev, using archival {\it XMM-Newton}, NICER, and {\it NuSTAR} observations. Joint spectral fitting of the NICER and {\it NuSTAR} data for \jsixt confirms the presence of a cyclotron resonant scattering feature (CRSF) at $\sim 50$~keV, implying a NS magnetic field strength of $\sim 6\times10^{12}$~G. We detect a cyclotron line at $\sim27$~keV in \jsevent, significantly higher than the previously reported energy of $\sim17$~keV, a difference that may reflect luminosity-dependent changes in the accretion flow. Using a double-Comptonization model, we identify the characteristic two-hump spectral shape in \jsixt and \jsevent at luminosities  $\lesssim10^{35}$~\lum, indicating that this spectral transition is a universal feature of accreting magnetized NSs independent of the source type. Furthermore, the ratio between the cyclotron line energy and the intersection energy of the two humps in both sources is consistent with the empirical relation established in Be/X-ray binaries, suggesting that the overheated polar cap atmosphere model used to explain X-ray pulsars in low-luminosity states is also applicable to SFXTs. In contrast, \jelev shows a smooth single-component spectrum at all flux levels, which indicates a relatively weak magnetic field of the NS. 
}


\maketitle
\nolinenumbers

\section{Introduction} 
\label{sec:intro}

High-mass X-ray binaries (HMXBs) are binary systems typically hosting a neutron star (NS), emitting X-rays powered by accreting matter  via stellar wind and/or Roche-lobe overflow from an early-type O or B companion star (for recent reviews, see, e.g., \citealt{Martinez-Nunez2017, Fornasini2024}). HMXBs can be classified as supergiant X-ray binaries (sgXBs) and Be/X-ray binaries (BeXBs) based on the spectral type of the companion star. In BeXBs, the NS orbits a Be-type companion surrounded by a circumstellar disk. These systems exhibit characteristic Type I outbursts and rare Type II outbursts, resulting in a wide range of X-ray luminosities. Supergiant fast X-ray transients (SFXTs), a subclass of wind-fed sgXBs, also exhibit extreme X-ray variability, with luminosities of $\sim10^{32}$~\lum\ \citep[e.g.,][]{intZand2005, Bozzo2010} during prolonged quiescent states and peak luminosities up to $10^{38}$~\lum\ \citep{Romano2015} during short (typically lasting a few hours) X-ray flares. Such pronounced luminosity variations in both BeXBs and SFXTs indicate substantial changes in the accretion process onto the NS, although the mechanisms responsible for their transient X-ray emission are not yet fully understood.

\begin{table*}
\centering
\renewcommand\arraystretch{1.3}
\caption{Summary of the \textit{XMM-Newton}, NICER and \textit{NuSTAR} observations of SFXTs sample used in this study.}
\begin{tabular}{ccccccc}
\hline
\hline
   Source & Observatory & ObsID & Start Time & Total Exposure & \\
 & &  & (UTC) & (ks) \\ \hline
  \jsixt &  NICER &  6752010101 & 2023-08-06T08:50:45 & 6.6 \\
   & NICER & 6752010102 & 2023-08-07T00:19:20  & 12.0  \\
  & {\it NuSTAR} & 30901022002 & 2023-08-06T09:16:08 & 79.1 \\ \hline
  \jsevent &  \textit{XMM-Newton} & 0744600101 & 2015-03-20T05:00:31  & 141.4 \\
   & {\it NuSTAR} & 90001005002 & 2015-03-20T00:51:07  & 79.2 \\ \hline
  \jelev &  \textit{XMM-Newton} &  0763210201 & 2016-02-14T21:07:53  & 23.9 \\
   & {\it NuSTAR} & 30101008002 & 2016-02-14T17:11:11  & 20.1 \\ \hline
\end{tabular}
\tablefoot{The exposure times for {\it NuSTAR} and {\it XMM-Newton} were obtained from FPMA and EPIC-PN instruments, respectively. } 
\label{table:obs}
\end{table*}

%

Recent observational studies show that the spectra of accreting X-ray pulsars (XRPs)  differ significantly at low and high mass accretion rates (see, e.g., \citealt{MushtukovTsygankov2024} for a review). At high luminosities ($L_{\rm X} \gtrsim 10^{36}$~\lum), the spectra of XRPs are generally described by an absorbed power law with an exponential cutoff at high energies \citep{Nagase1989,2005AstL...31..729F}.
However, at lower luminosities ($L_{\rm X} \lesssim 10^{35}$~\lum), the spectra of some XRPs exhibit a characteristic ``two-hump'' shape \citep[e.g.,][]{Tsygankov2019a,Tsygankov2019b}. In such systems, where the NS is highly magnetized ($B\sim10^{12}$--$10^{13}~$G), the accreted matter near the magnetosphere couples to magnetic field lines and is channeled onto the polar caps of the NS, 
forming hot spots.
The characteristic two-hump spectral shape can be explained by radiation emerging from an accretion-heated atmosphere of a strongly magnetized NS \citep{Mushtukov2021,Sokolova-Lapa2021}.
The soft component is associated with thermal radiation reprocessed in the atmosphere, while the hard component is shaped by cyclotron emission and resonant Compton scattering.
Cyclotron photons produced by collisional excitation of electrons undergo resonant scattering during their propagation through the atmosphere, which broadens the high-energy spectral component. 
Both theoretical studies mentioned above explore this general physical picture using different approximations and numerical approaches to radiative transfer.

\cite{Zalot2026} performed a systematic analysis of BeXBs in quiescence and discovered a linear correlation between the intersection energy of the two humps ($E_{\rm int}$) and the energy of cyclotron line ($E_{\rm cyc}$). They empirically determined the ratio to be $E_{\rm cyc} \approx 4.15 E_{\rm int}$, establishing a quantitative link between the spectral morphology and strength of the NS magnetic field, and proposed this relation as a possible way to estimate the magnetic field strength when cyclotron lines are not directly detectable. Recently, \cite{Ducci2026} investigated the spectral properties of IGR~J00370+6122 at luminosities as low as \(\sim 10^{34}~\mathrm{erg\,s^{-1}}\) and reported this characteristic two-hump shape in an SFXT for the first time, revealing that SFXTs may share similar physical conditions in the emission region as BeXBs at low luminosities. However, it remains unclear whether the $E_{\rm int}-E_{\rm cyc}$ relation also applies to SFXTs. Given that cyclotron lines have been detected in only a few SFXTs, partly due to their typically low luminosities, this relation could provide an alternative method for estimating the NS magnetic field strength in these systems.

\jsixt, \jsevent, and \jelev are among the few SFXTs for which both NS spin periods and cyclotron resonance scattering features (CRSFs) have been reported, suggesting that these systems host strongly magnetized NSs. In this paper, we utilized data from the {\it NuSTAR}, {\it XMM-Newton}, and NICER observatories to study the spectral properties of these sources during their different flux states and to test the $E_{\rm int}-E_{\rm cyc}$ relation.

\section{Sample of SFXTs}

\jsixt is an 
SFXT that was discovered by {\it INTEGRAL} during the observations of 4U 1630$-$47 \citep{Tomsick2004}. It consists of a NS with a spin period of $\sim1210$~s \citep{Sidoli2012} and a BN0.5\,Ia supergiant companion \citep{Coleiro2013}. It is an eclipsing binary system with an orbital period of $\sim3.74$~d \citep{Coley2015}, located at a distance of 4.9~kpc \citep{Neumann2023}. The quasi-spherical accretion model suggests a NS magnetic field strength of $\sim10^{14}$~G  \citep{Drave2013b}, whereas the cyclotron line detected at $\sim50$~keV implies a field strength of $\sim5\times10^{12}$~G \citep{Islam2023, Natalucci2024, Devaraj2026}. Superorbital modulations have been discovered in this source with a modulation period of $\sim14.7$~d  \citep{Drave2013a}. \cite{Islam2014} found no significant differences in the timing and spectral properties between the superorbital maximum and minimum phases. \citet{Sguera2006} and \citet{Ducci2010} reported strong X-ray flux variations spanning two orders of magnitude on timescales of about an hour, with peak luminosities reaching $\sim10^{37}~\rm erg\,s^{-1}$ during flares \citep[e.g.,][]{Romano2012}.
\jsevent is the prototypical SFXT discovered by {\it INTEGRAL} on 2003 September 17 \citep{Sunyaev2003}. The optical companion was identified as an early O9\,I supergiant \citep{Gimenez-Garcia2016} at a distance of $2.4\pm0.5$~kpc \citep{Neumann2023}. \citet{Bhalerao2015} reported a cyclotron line at 17~keV, suggesting that the compact object is a NS with a magnetic field of $B\simeq 1.5 \times 10^{12}$~G. A possible pulsation signal with a period of $\sim$71~s \citep{Drave2012} (as well as a hint of 11.6~s; \citealt{Romano2015}) was detected, which may correspond to the spin period of the NS. However, \cite{Bozzo2016} did not detect either of these periodic signals in their {\it NuSTAR} and {\it XMM-Newton} observations.  The orbital period was estimated to be $4.926\pm0.001$~d \citep{Clark2009}. The luminosity of \jsevent was observed to span a wide range, from $\sim$$10^{32}$~\lum\ in quiescence \citep{intZand2005} to $\sim$$10^{38}$~\lum\ during flares \citep[e.g.,][]{Pellizza2006}.
\jelev was discovered in 2005 with the {\it INTEGRAL} satellite \citep{Lubinski2005}. The NS has a spin period of $\sim$187~s \citep{Swank2007}. It orbits a B0.5 Ia companion HD 306414 \citep{Negueruela2005} in an eccentric orbit ($e\gtrsim0.8$) with an orbital period of $\sim164.6$~d \citep{Sidoli2007, Romano2009}, and is unique among SFXTs in undergoing strictly periodic X-ray outbursts. It is located at  approximately 6.5~kpc based on Gaia DR3 measurements \citep{Neumann2023}. For the NS, an earlier observation hinted at a CRSF at $\sim$17~keV, corresponding to a magnetic field of $2\times10^{12}$~G \citep{Sidoli2017}, while a  recent analysis of pulse profile evolution strongly favor a lower field of $\sim10^{11}$~G under the quasi-spherical settling accretion model \citep{Sidoli2020}.

\section{Observations and data reduction} 
\label{sec:data}

To better constrain the spectra of these sources, we selected archival data providing simultaneous observations of {\it XMM-Newton} (or NICER) and {\it NuSTAR}, and covering a wide energy range from 0.5~keV to 79~keV. This allowed us to reveal the expected two-hump spectral shape at low luminosities over a wider energy band. The details of the observations are presented in Table~\ref{table:obs}.
\subsection{XMM-Newton}
The \textit{X-ray Multi-Mirror Mission} ({\it XMM-Newton}) is an X-ray observatory equipped with the European Photon Imaging Camera (EPIC), consisting of two MOS and one PN CCD cameras and covering the 0.3--12~keV band. The {\it XMM-Newton} data were reduced and analyzed using the Science Analysis System ({\sc sas} v22.1.0)\footnote{\url{https://www.cosmos.esa.int/web/xmm-newton/sas}} software with the latest calibration datasets. We used the {\tt epproc} and {\tt emproc} tasks to produce calibrated and concatenated event lists for the PN and MOS cameras, respectively. Since our SFXT sources are highly variable during the observations, an annular region was used to avoid pile-up issues. Specifically, the source region was identified as a circular region with a 30\arcsec\ radius at low count-rate phases, and as an annulus with an inner radius of 15\arcsec\ and an outer radius of 30\arcsec\ at high count-rate phases. The background was extracted from a nearby, source-free circular region with a 60" radius on the same CCD. Single and double events (PATTERN $\leq$ 4) were extracted for the PN camera and events between single and quadruple (PATTERN $\leq$ 12) were selected for the MOS cameras. The response matrices and ancillary response files were generated using the {\tt rmfgen} and {\tt arfgen} tasks, respectively, and the net light curves were obtained using the {\tt epiclccorr} task. We used the {\tt gtibuild} tool to generate good time intervals (GTIs).
\subsection{NICER}
Neutron star Interior Composition Explorer (NICER; \citealt{Gendreau2016}) is a payload installed on the International Space Station, dedicated to timing and spectral study in 0.2--12 keV energy range. The data were reduced and analyzed using the official software \textsc{heasoft} v6.36\footnote{\url{https://heasarc.gsfc.nasa.gov/docs/software/lheasoft}}/\textsc{nicerdas} package (version {2025-03-11\_V013a), along with the latest calibration database (CALDB) version  20240206. 
The clean event files were produced with {\tt nicerl2}  with the filter criteria {NICERV6}. We combined the data from two observations using the {\tt niobsmerge} task. Light curves and spectra were extracted using {\tt nicerl3-lc} and {\tt nicerl3-spect} tools, respectively. The tool {\tt nimaketime} was used to generate specific GTIs. The background was estimated by using {\sc scorpeon}\footnote{\url{https://heasarc.gsfc.nasa.gov/docs/nicer/analysis_threads/scorpeon-overview}} model. The count rates were normalized to 52-FPM equivalent intensity. 
\subsection{NuSTAR}
Launched on 2012 June 13, the \textit{Nuclear Spectroscopic Telescope Array} (\textit{NuSTAR}; \citealt{Harrison2013}) is a hard X-ray telescope operating in a wide energy band of 3--79~keV. It consists of two co-aligned hard X-ray grazing incidence telescopes, each telescope having its own focal plane module, FPMA and FPMB. The \textit{NuSTAR} data were reduced and analyzed using the standard data analysis software \textsc{nustardas} v2.1.5. The latest CALDB (version 20260923) was used during the data reduction. We used the task \texttt{nupipeline} to filter the clean event files. Source region was identified as a circular region with a radius of 30\arcsec\ and the background area was selected within an annular region centered on the source with inner and outer radii of 60\arcsec\ and 100\arcsec, respectively. The source spectra, background spectra, response matrices, ancillary matrices, and light curves were created using the \texttt{nuproducts} task. Spectra during the user-defined GTIs were extracted by setting the parameter \texttt{usrgtifile} of \texttt{nuproducts}. 
We rebinned the spectra with the \texttt{grppha} tool to ensure a minimum of 25 counts per energy bin. The spectral analysis was performed using \textsc{xspec} version 12.15.1 \citep{Arn96}. All uncertainties in this paper correspond to a confidence level of 90\%, estimated by running Monte Carlo Markov Chains with a length of 20,000 with a burn-in of 10,000 using the Goodman-Weare algorithm.

\begin{table*}
\centering
\caption{Flux-averaged and flux-resolved best-fit spectral parameters of \jsixt\ using the 2023 NICER and {\it NuSTAR} observations.}
\label{tab:j16418_spec}
\renewcommand{\arraystretch}{1.3}
\begin{tabular}{lccc}
\hline
\hline
Parameter & Average state & Low state & High state \\
\hline
$C_{\rm FPMB}$ & $1.02\pm0.01$ & $1.07\pm0.01$ & $1.02\pm0.01$ \\
$C_{\rm NICER}$ & $1.06\pm0.02$ & $1.00\pm0.04$ & $0.83_{-0.02}^{+0.03}$ \\
$N_{\rm H}$ ($10^{22}$ cm$^{-2}$) & $3.4_{-0.1}^{+0.2}$ & $1.0_{-0.5}^{+0.4}$ & $4.0_{-0.3}^{+0.4}$ \\
$T_{0}$ (keV) & $1.05_{-0.02}^{+0.01}$ & $1.31\pm0.01$ & $0.97_{-0.05}^{+0.04}$ \\
$kT_{\rm e, low}$ (keV) & $8.3\pm0.3$ & $5.6_{-1.0}^{+0.8}$ & $7.9\pm0.2$ \\
$\tau_{\rm low}$ & $10.2_{-0.2}^{+0.3}$ & $6.6_{-0.8}^{+1.0}$ & $11.0\pm0.2$ \\
Norm$_{\rm low}$ ($10^{-3}$) & $2.55_{-0.07}^{+0.10}$ & $0.38_{-0.06}^{+0.07}$ & $5.3\pm0.2$ \\
$kT_{\rm e, hard}$ (keV) & -- & $7.5\pm0.2$ & -- \\
$\tau_{\rm hard}$ & -- & $100.0$ (fixed) & -- \\
Norm$_{\rm hard}$ ($10^{-4}$) & -- & $0.91_{-0.03}^{+0.04}$ & -- \\

$E_{\rm gau}$ (keV) & $6.31_{-0.04}^{+0.06}$ & $6.48\pm0.01$ & $6.22_{-0.08}^{+0.07}$ \\
$\sigma_{\rm gau}$ (keV) & $0.1$ (fixed) & $0.1$ (fixed) & $0.1$ (fixed) \\
Norm$_{\rm gau}$ ($10^{-4}$) & $0.47\pm0.10$ & $0.20\pm0.01$ & $1.1\pm0.3$ \\

$E_{\rm cyc}$ (keV) & $53.9_{-1.8}^{+2.3}$ & $53.9$ (fixed) & $53.9$ (fixed) \\
$\sigma_{\rm cyc}$ (keV) & $9.3_{-1.0}^{+0.9}$ & $9.4$ (fixed) & $9.4$ (fixed) \\
$d_{\rm cyc}$ & $1.4\pm0.4$ & $1.4$ (fixed) & $1.4$ (fixed) \\
$L_{\rm 0.5-60~keV}$ ($10^{34}$~\lum) & $85.2\pm0.6$ & $9.9^{+0.1}_{-0.2}$ & $146.0^{+1.8}_{-0.6}$  \\
\hline
$\chi^2$ (d.o.f) & $1664.8$ ($1627$) & $384.8$ ($336$) & $1189.9$ ($1182$) \\
\hline
\end{tabular}
\tablefoot{$C_{\rm FPMB}$ and $C_{\rm NICER}$ are the cross-calibration constants for {\it NuSTAR}/FPMB and NICER relative to {\it NuSTAR}/FPMA (which is fixed at 1). $N_{\rm H}$ is the equivalent hydrogen column density of the interstellar absorption. $T_0$ is the input seed photon temperature for the Comptonization model. $kT_{\rm e}$ and $\tau$ are the electron temperature and optical depth of the Comptonized plasma, respectively, where the subscripts `low' and `hard' denote the soft and hard {\tt comptt} components. $E_{\rm gau}$, $\sigma_{\rm gau}$, and $Norm_{\rm gau}$ represent the centroid energy, width, and normalization of the Fe K$\alpha$ emission line. $E_{\rm cyc}$, $\sigma_{\rm cyc}$, and $d_{\rm cyc}$ correspond to the centroid energy, width, and optical depth of the cyclotron line. The luminosities ($L_{\rm 0.5-60~keV}$) are unabsorbed values in the 0.5--60~keV band, calculated by assuming isotropic source at a distance of 4.9~kpc.}
\end{table*}

\section{Results} 
\label{sec:res}

\subsection{\jsixt}
\label{sec:j16418}
Using different phenomenological and physical models, \cite{Devaraj2026} described the spectrum of \jsixt observed by {\it NuSTAR} in 2023 and confirmed the cyclotron line at $\sim$50~keV. We performed a flux-averaged spectral analysis using the same data set but with the addition of data from NICER to further constrain the spectral parameters. We fitted the spectrum in the 1--79~keV band with a thermal Comptonization model {\tt comptt} \citep{Titarchuk1994}. A {\tt constant} component was used to offset differences in calibration between instruments and a {\tt tbabs} component was used to describe the interstellar photoelectric absorption. The element abundances were set to {\tt wilms} \citep{Wilms2000}, along with the photoelectric cross-sections of {\tt vern}. Iron K$\alpha$ line is seen in the spectrum at $\sim$6.4~keV, which was modeled by a {\tt gaussian} component and the width of the line was fixed at 0.1~keV.

A significant negative residuals appear around 50~keV, corresponding to $\chi^2=1977.8$ for 1630 d.o.f. In order to improve the fit,  we introduced a {\tt gabs} component to the model. In this case, we obtained acceptable fitting results with $\chi^2$ (d.o.f.) of 1664.8 (1627). The energy of the cyclotron line is estimated to be $E_{\rm cyc}=53.9^{+2.3}_{-1.8}$~keV, consistent with previous reports \citep{Devaraj2026}. The spectral fitting results and the best-fit parameters of the flux-averaged spectra of \jsixt are shown in Fig.\ref{fig:avg_spec_j16418} and Table~\ref{tab:j16418_spec}, respectively.

\begin{figure}
\centering
\includegraphics[width=0.99\linewidth]{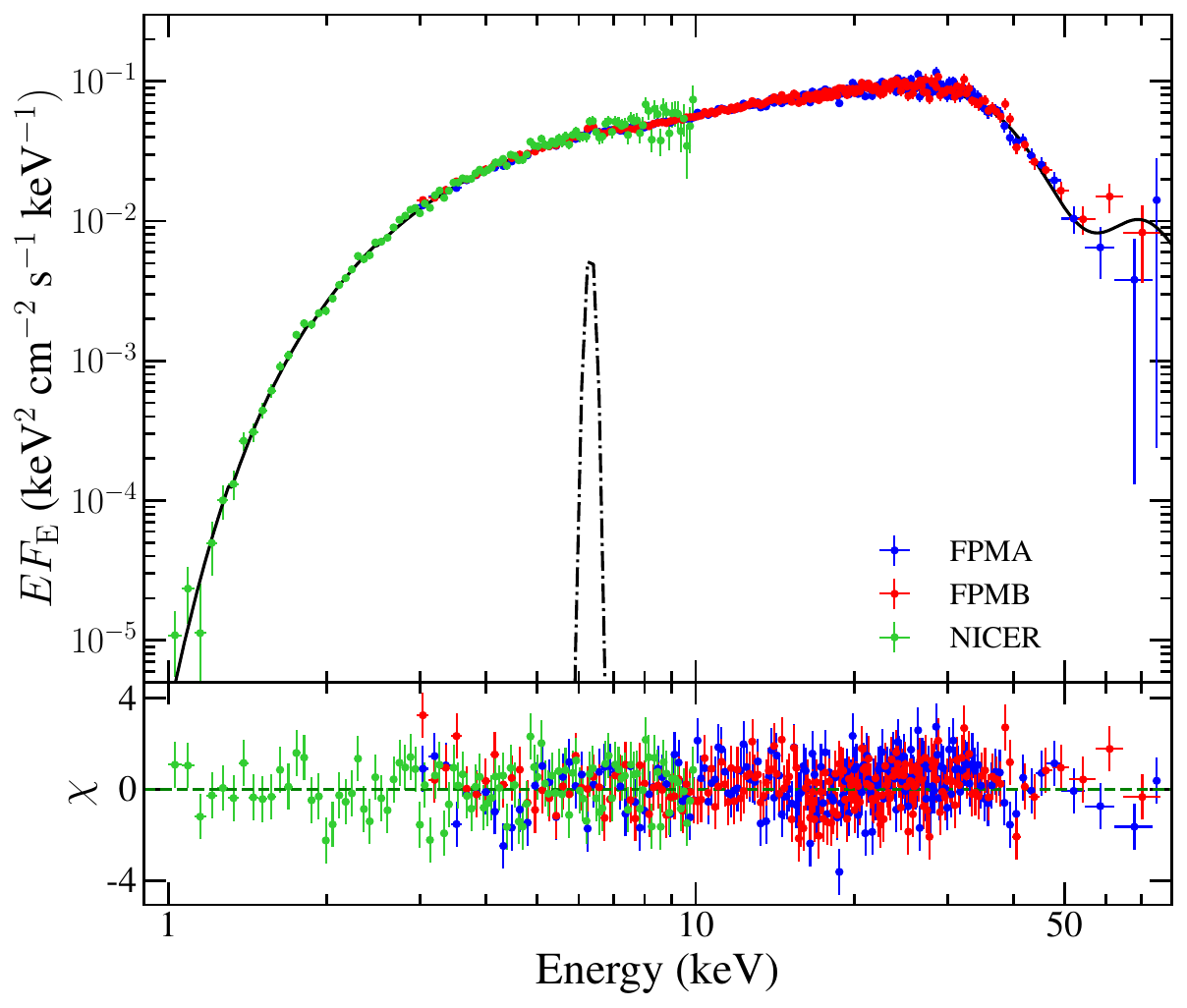}

\caption{Flux-averaged spectrum of \jsixt obtained from NICER and {\it NuSTAR} observations. The solid line represents the best-fit spectral model {\tt const$\times$tbabs$\times$(comptt+gauss)$\times$gabs} and the dash-dotted line shows the Fe~K$\alpha$ line. The corresponding residuals are displayed in the bottom panel.}
\label{fig:avg_spec_j16418}
\end{figure}

In Fig.~\ref{fig:lc_j16418}, we present the light curves of \jsixt from the NICER and \textit{NuSTAR} observations. To investigate the spectral properties of \jsixt in different luminosity states, particularly in quiescence, we extracted the spectra according to the count rate. For {\it NuSTAR}, we extracted the spectra when the source flux was below 1.5~count~s$^{-1}$ and above 5~count~s$^{-1}$ to probe the low and high states, respectively (the green lines in Fig.~\ref{fig:lc_j16418}). To determine the corresponding NICER count-rate thresholds, we performed spectral fitting of the {\it NuSTAR} observations at different intensity levels. The resulting spectral parameters were then used as input to WebPIMMS\footnote{\url{https://heasarc.gsfc.nasa.gov/cgi-bin/Tools/w3pimms/w3pimms.pl}} to simulate the expected NICER count rates. With this approach, we obtain a lower threshold of 2.5~count~s$^{-1}$ and an upper threshold of 8~count~s$^{-1}$. 
%

\begin{figure}
\centering
\includegraphics[width=0.95\linewidth]{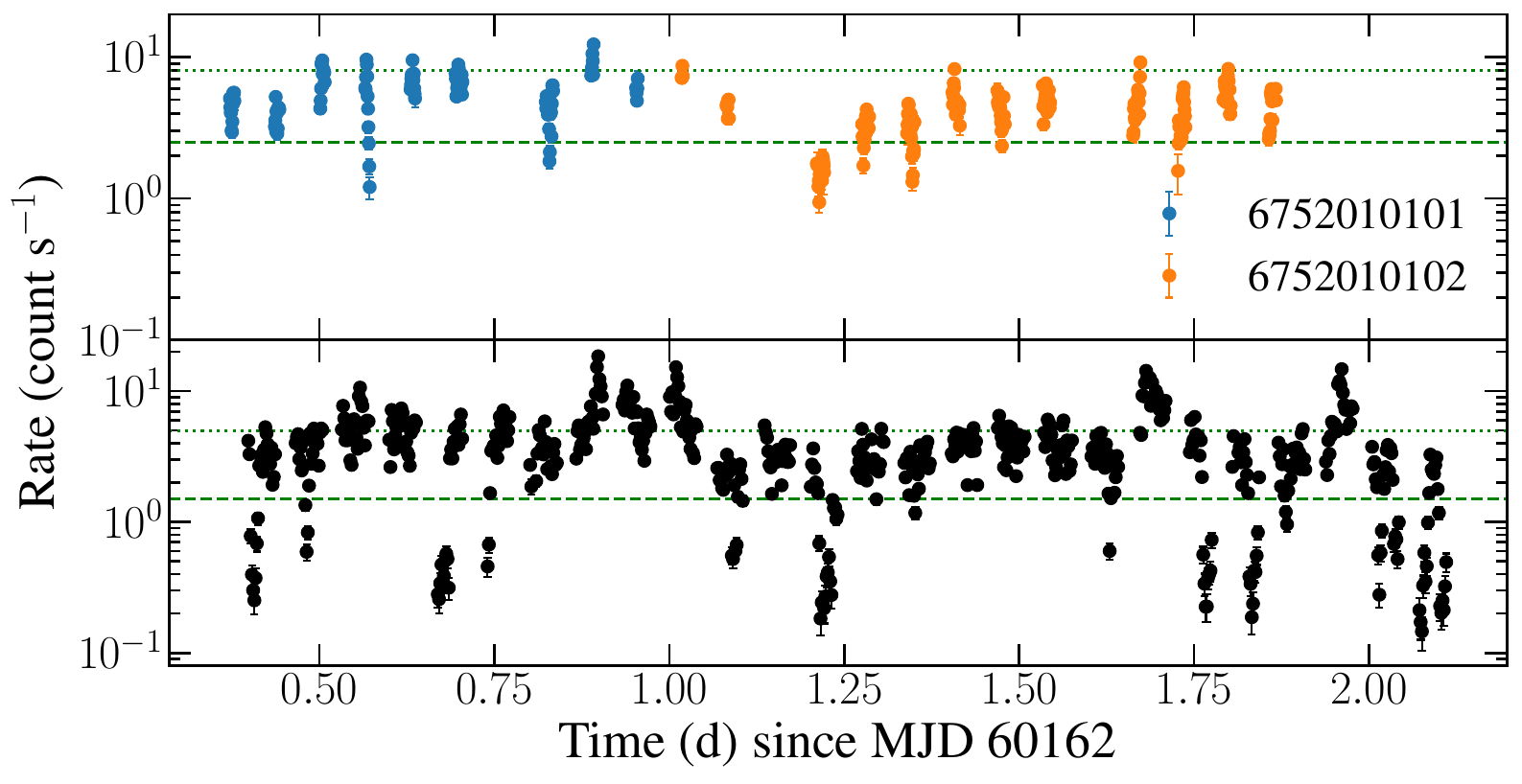} 

\caption{Background-subtracted light curves of \jsixt observed by NICER (top; 0.5--10~keV) and {\it NuSTAR}/FPMA (bottom; 3--79~keV). The green dashed and dotted lines indicate the count-rate thresholds (2.5 and 8.0~count~s$^{-1}$ for NICER; 1.5 and 5.0~count~s$^{-1}$ for {\it NuSTAR}), defining the low and high flux states used for the flux-resolved spectroscopy.} 
\label{fig:lc_j16418}
\end{figure}

%

%
We first used a single Comptonization model {\tt const$\times$tbabs$\times$comptt$\times$gabs} to describe these two spectra, with the parameters of cyclotron line fixed at their flux-averaged values (see Table~\ref{tab:j16418_spec}) due to the low statistics. This model yields a satisfactory fit for the high state with $\chi^2$ (d.o.f.) = 1189.9 (1182). However, the fit for the low state with $\chi^2$ (d.o.f.) = 397.9 (338) is not acceptable, with systematic residuals clearly visible above $\sim20$~keV. We attempted to improve the fit by either allowing the {\tt gabs} parameters to vary or fixing its depth at zero, but neither approach yielded a significant improvement. This may indicate a transition to the two-hump spectral shape, typically observed in accreting XRPs when the luminosity drops to $\lesssim 10^{35}$~\lum. To account for this, we applied a two-component Comptonization model ({\tt comptt + comptt}) with the seed photon temperatures ($kT_0$) of the two components tied together. This model has been widely used for XRPs in quiescence \citep[e.g.,][]{Doroshenko2012, Tsygankov2019b}. Since the electron temperature ($kT_{\rm e}$) and optical depth ($\tau$) are strongly degenerate, we fixed the optical depth of the high energy component at $\tau_{\rm hard} = 100$ to achieve an acceptable fit, following the same approach as in previous studies \citep[e.g.,][]{Salganik2025}. This two-component model provided a significantly improved fit for the spectra yielding $\chi^2$ (d.o.f.) = 384.7 (336). In Fig.~\ref{fig:j16418_flux_resoved_spec}, we present the spectral fitting results for \jsixt in different flux states. We used the {\tt simftest} tool in \textsc{xspec} with 10,000 simulations to calculate the significance of this additional component, indicating that the {\tt comptt + comptt} model is statistically favored over the single {\tt comptt} model with a p-value of $8\times10^{-3}$. The luminosity of the low state was estimated to be $L_{\rm 0.5-60~keV}=(9.9^{+0.1}_{-0.2})\times10^{34}$~\lum. This luminosity is comparable to the typical value at which two-hump spectra are observed in BeXBs \citep{Zalot2026}. To estimate the intersection energy of the two spectral components and examine the relation $E_{\rm int}$--$E_{\rm cyc}$ compared to that observed in BeXBs, we fitted the low-state spectrum of \jsixt using the phenomenological model proposed by \cite{Zalot2026}. This model yields an intersection energy $E_{\rm int}$ of the two components of $10.6^{+1.2}_{-0.9}$~keV. Using the cyclotron line energy of $E_{\rm cyc} = 53.9^{+2.3}_{-1.8}$~keV, we obtain a ratio of $E_{\rm cyc}/E_{\rm int} = 5.1^{+0.5}_{-0.6}$. The best-fit spectral parameters for the high and low states are presented in Table~\ref{tab:j16418_spec}.

\begin{figure}
\centering
\includegraphics[width=0.99\linewidth]{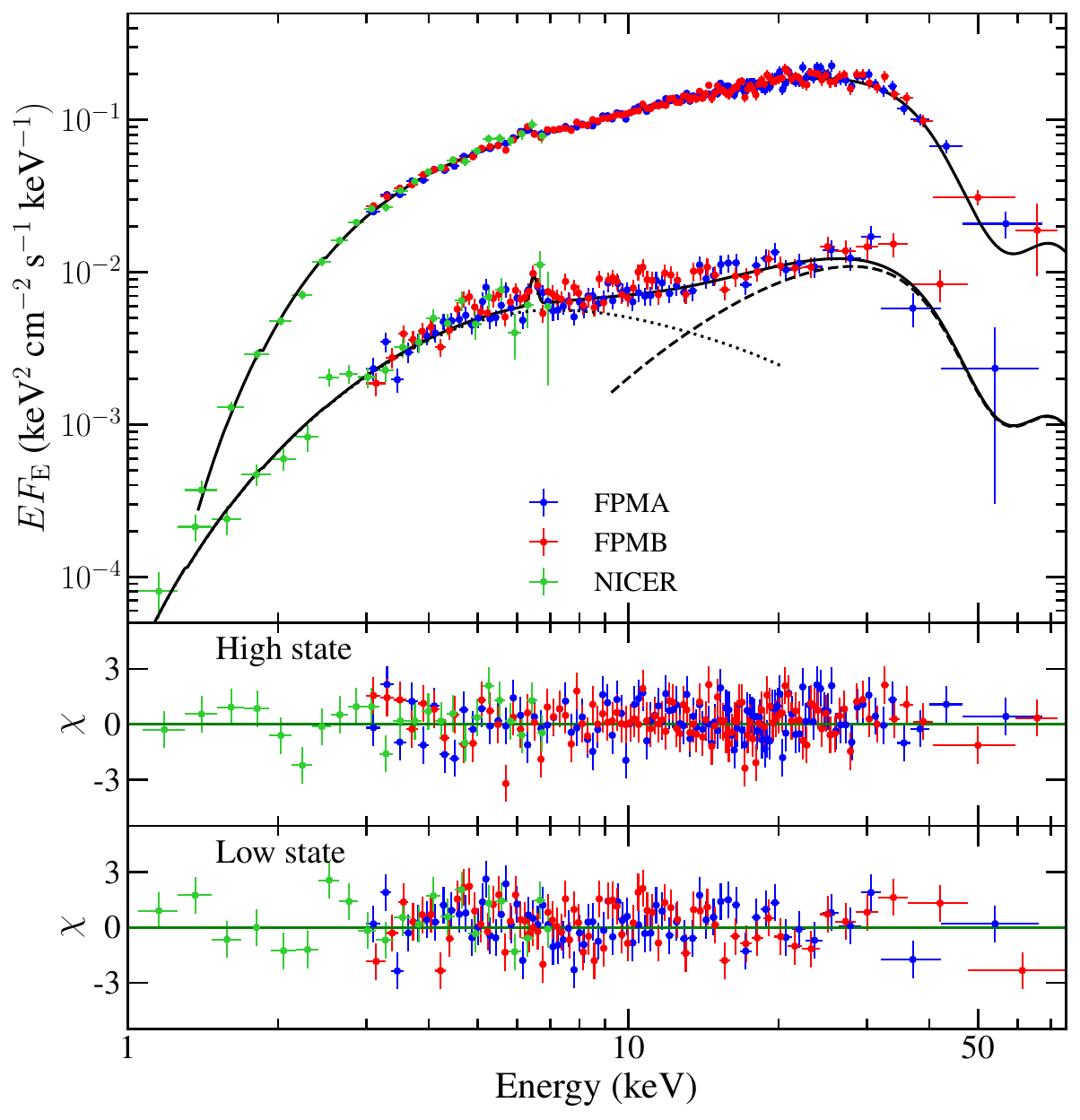}
\caption{Unfolded spectra of \jsixt for the low ($9.9\times10^{34}$~\lum) and high ($1.5\times10^{36}$~\lum) states, respectively. The fitted models are {\tt const$\times$tbabs$\times$(comptt+comptt+gauss)$\times$gabs} for the low state and {\tt const$\times$tbabs$\times$(comptt+gauss)$\times$gabs} for the high state. The dotted and dashed lines represent the low and high energy {\tt comptt} components, respectively. Corresponding residuals are displayed in the lower panels. }
\label{fig:j16418_flux_resoved_spec}
\end{figure}


\subsection{\jsevent}
\label{sec:j17544}

For \jsevent, the {\it XMM-Newton} and {\it NuSTAR} light curves  show that the source first was found in a quiescent phase lasting approximately 120~ks, followed by a dramatic X-ray flare with the flux varying by more than three orders of magnitude (see Fig.~\ref{fig:lc_j17544}). Therefore, to search for cyclotron lines, we performed a spectral analysis using data from the flaring interval (the green shaded region in Fig.~\ref{fig:lc_j17544}). \cite{Bhalerao2015} detected a cyclotron line at $\sim17$~keV and its harmonic at $\sim$34~keV. However, \cite{Bozzo2016} found no absorption features in the spectrum using neither the {\tt bbodyrad+po$\times$highecut} model nor the bulk and thermal Comptonization ({\tt BWcycl}) model \citep{Ferrigno2009}. \citet{Bozzo2016} noted that the spectrum of \jsevent exhibits a bimodal shape and the ``valley'' in the energy range 6--9~keV was interpreted as the intersection of the soft and hard components. Consistent with the treatment of \jsixt in Sect.~\ref{sec:j16418}, here we used a double-Comptonization model to describe the spectrum. 
%

\begin{figure}
\centering
\includegraphics[width=0.95\linewidth]{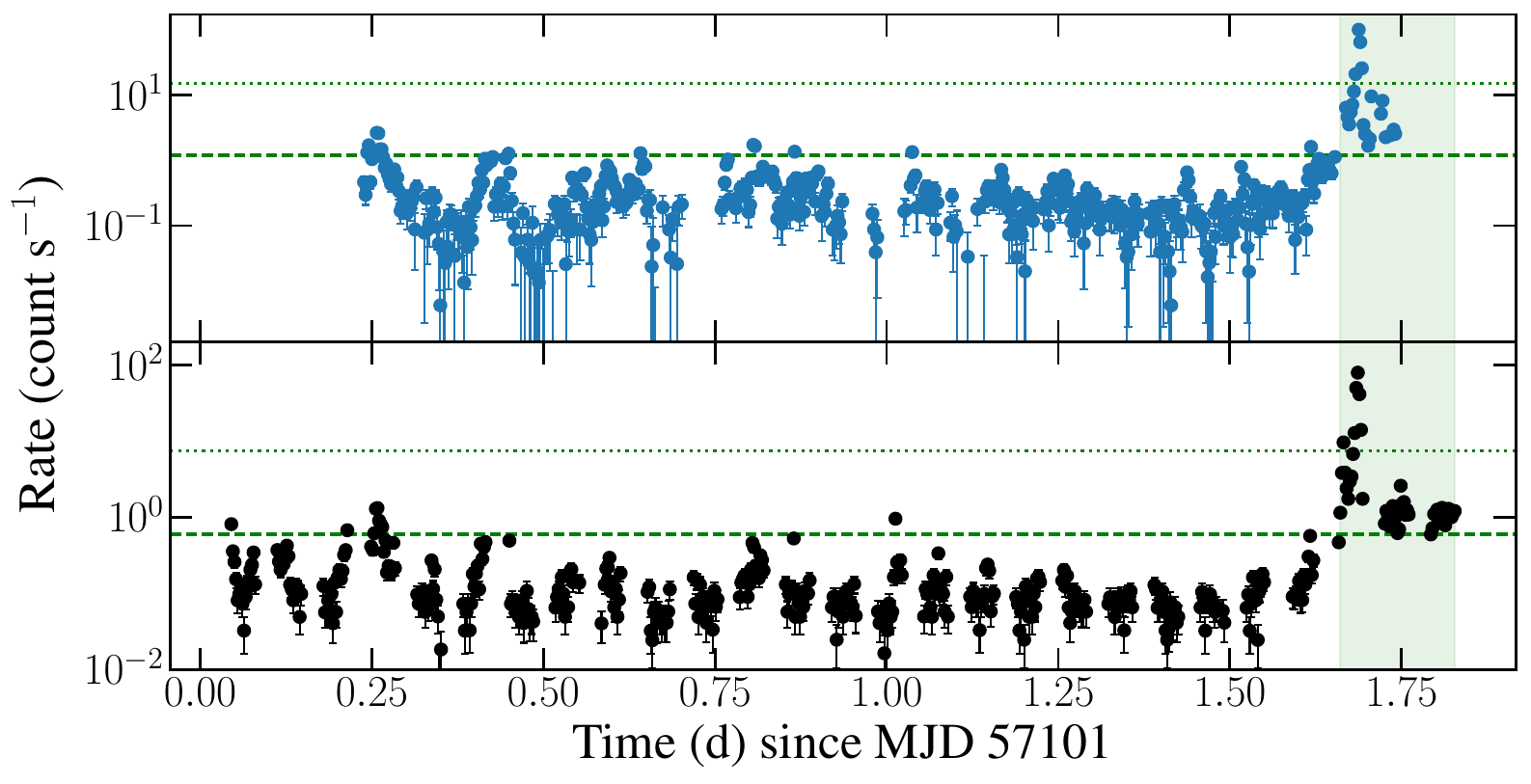} 

\caption{Background-subtracted light curves of \jsevent observed by {\it XMM-Newton}/EPIC-PN (top; 0.3--10~keV) and {\it NuSTAR}/FPMA (bottom; 3--79~keV). The green shaded region indicates the flaring interval targeted for a dedicated search for cyclotron lines. The green dashed and dotted lines indicate the count-rate thresholds (1.2 and 15~count~s$^{-1}$ for {\it XMM-Newton}; 0.6 and 7.5~count~s$^{-1}$ for {\it NuSTAR}), separating the low, middle, and high states for the flux-resolved spectroscopy.}
\label{fig:lc_j17544}
\end{figure}

\begin{figure}
\centering
\includegraphics[width=0.95\linewidth]{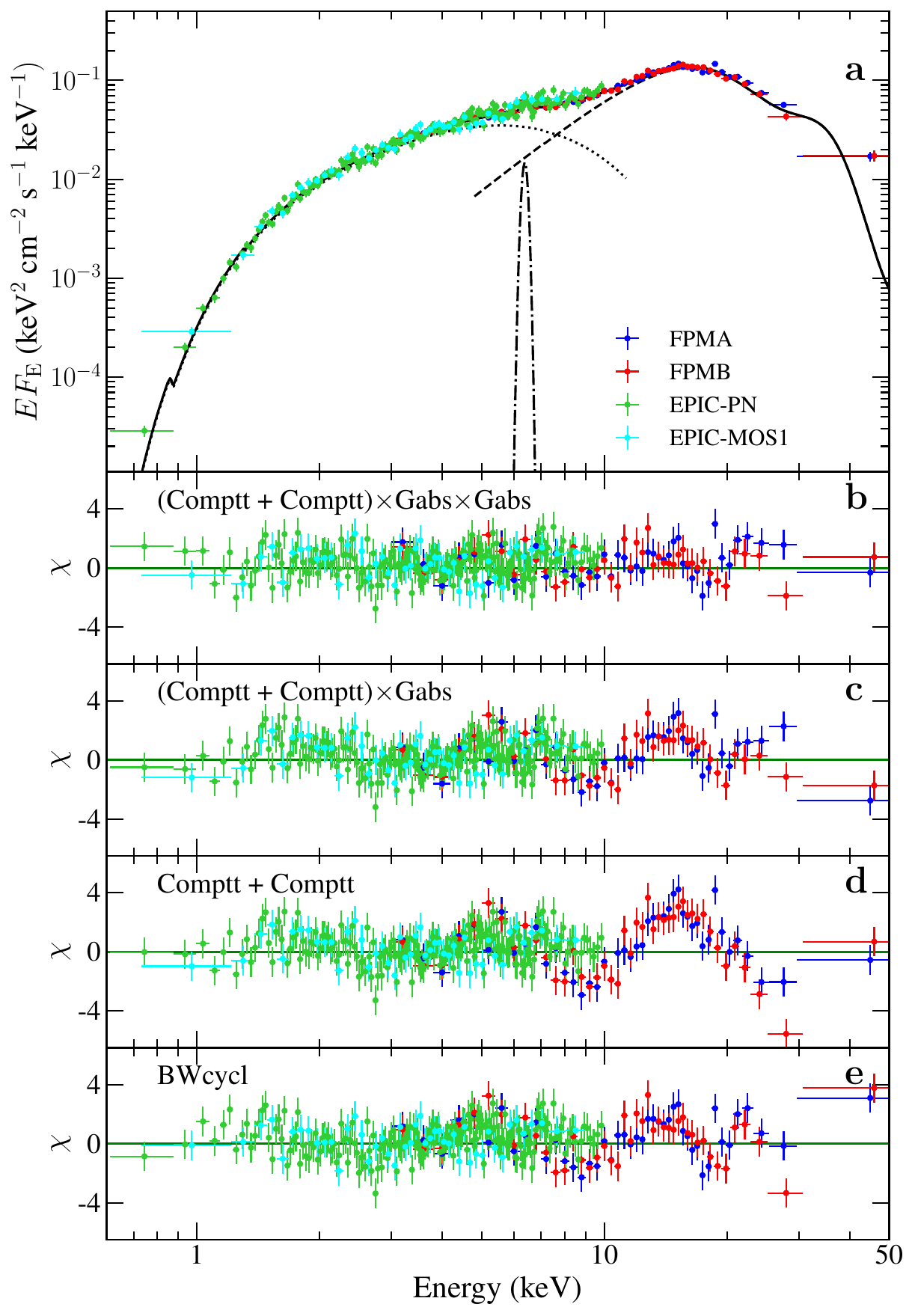}
\caption{Unfolded spectrum of \jsevent during the flare (see green shaded region in Fig.~\ref{fig:lc_j17544}), obtained from the {\it XMM-Newton} and {\it NuSTAR} observations. The solid line represents the best-fit model of {\tt const$\times$tbabs$\times$(comptt+comptt+gauss)$\times$gabs$\times$gabs}, the dotted  and dashed lines represent the low and hard spectral components, respectively. The lower panels show the residuals of spectra fitted by a double-Comptonization model and the same model with different prescriptions for the cyclotron absorption line. The bottom panel represents the residuals obtained by fitting the spectrum using the {\tt BWcyc} model.}
\label{fig:avg_spec_J17544}
\end{figure}

\begin{figure}
\centering
\includegraphics[width=0.95\linewidth]{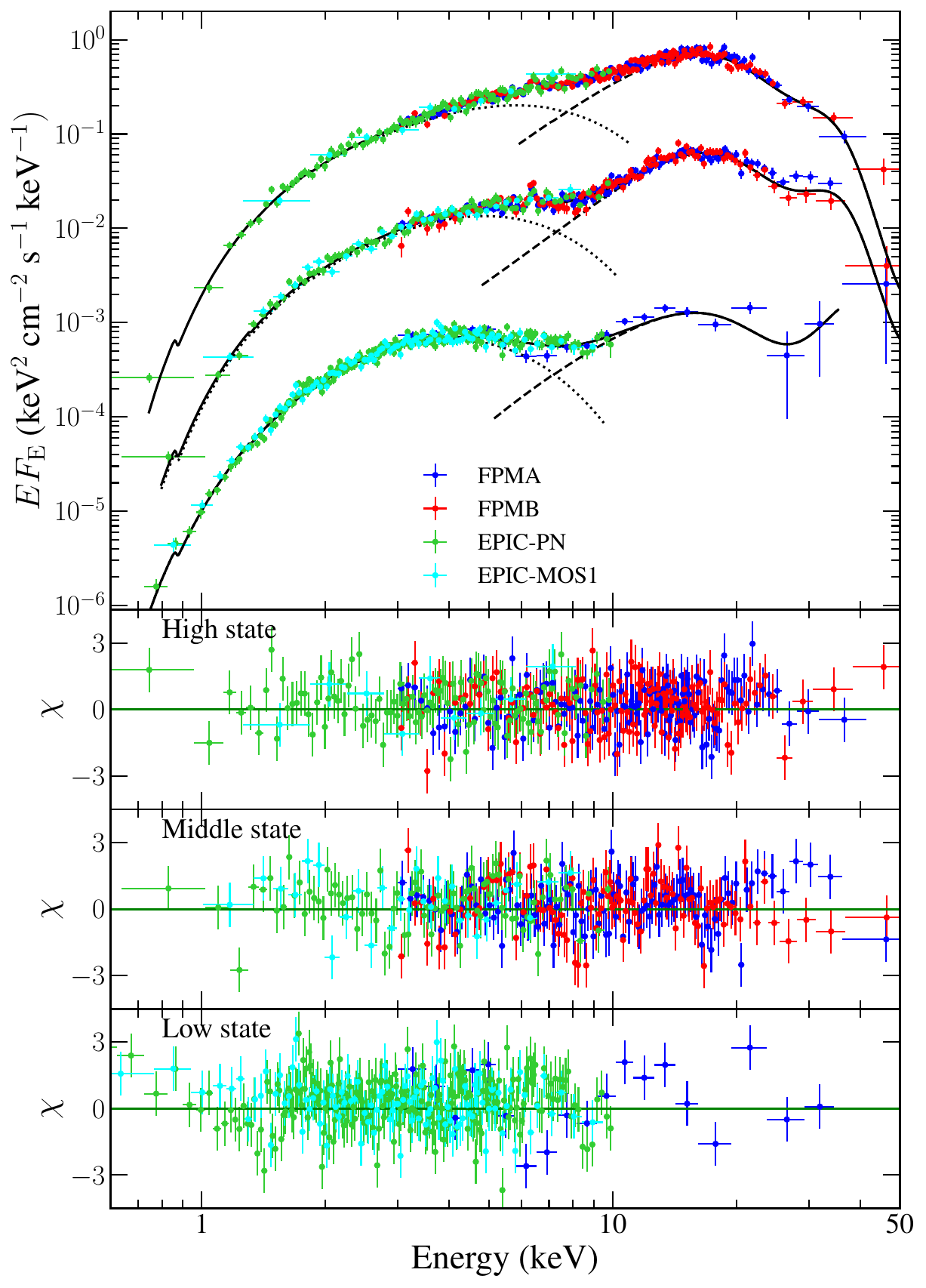}
\caption{Unfolded spectra of \jsevent, extracted from different states with luminosities of $1.1\times10^{36}$~\lum\ (high state), $9.1\times10^{34}$~\lum\ (middle state) and $3.4\times10^{33}$~\lum\ (low state), respectively. The solid lines represent the best-fit spectral model, i.e., {\tt const$\times$tbabs$\times$(comptt+comptt)$\times$gabs$\times$gabs}, the dotted and dashed lines represent the low and high component, respectively. For the low state, we exclude the {\it NuSTAR}/FPMB spectrum since the source region was contaminated by stray light. }
\label{fig:j17544_flux_resoved_spec}
\end{figure}

%
We first fitted the spectrum during the flaring state using a double-Comptonization model. However, the fitting results show significant deviations between the observed spectrum and the model predictions at energies above 10~keV (see Fig.~\ref{fig:avg_spec_J17544}d), resulting in $\chi^2$ (d.o.f.) of 1943.4 (1608). We added a {\tt gabs} component to the model, which slightly improved fit with a cyclotron line energy of $\sim$26~keV. However, a negative residual is still present at $>$40~keV, likely caused by its harmonic (Fig.~\ref{fig:avg_spec_J17544}c). We therefore added an additional {\tt gabs} component to the model, yielding best-fit energies of $\sim$26 and 51~keV for the fundamental cyclotron line and its first harmonic, respectively. Since these energies are in a 1:2 ratio within uncertainties and our goal is to estimate the fundamental energy, we fixed the harmonic energy at twice the fundamental energy to improve the stability of the fit. The {\tt const$\times$tbabs$\times$(comptt+comptt+gauss)$\times$gabs$\times$gabs} model provided an acceptable fit to the spectrum, yielding $\chi^2$/d.o.f.=1653.9/1604 (Fig.~\ref{fig:avg_spec_J17544}b). The best-fit cyclotron line energy $E_{\rm cyc}=26.6^{+1.2}_{-0.6}$~keV is higher than the $\sim$17~keV reported by \cite{Bhalerao2015}. Following \cite{Bozzo2016}, we tried to use a bulk and thermal Comptonization {\tt BWcycl} model to fit the spectrum, yet negative residuals are still clearly observable around 26~keV (Fig.~\ref{fig:avg_spec_J17544}e), indicating that the detection of this feature does not dependent on the choice of the continuum model.
%

%
For the flux-resolved spectral analysis, we divided the observations into low, middle, and high states based on the intensity thresholds indicated by the two green lines in Fig.~\ref{fig:lc_j17544}. These states correspond to {\it NuSTAR} count rates of $<$0.6, 0.6--7.5, and $>$7.5~count~s$^{-1}$, respectively. Using the same procedure described in Sect.~\ref{sec:j16418}, the corresponding {\it XMM-Newton} count-rate thresholds were determined to be 1.2 and 15~count~s$^{-1}$.

For consistency, we adopted the same {\tt const$\times$tbabs$\times$(comptt + comptt)$\times$gabs$\times$gabs} model throughout the spectral analysis of all states. To facilitate spectral fitting, the parameters of the two {\tt gabs} components were fixed to their flare-state values, except for their depths ($d_{\rm cyc, 1}$ and $d_{\rm cyc, 2}$). \citet{Salganik2026} reported the luminosity-dependent evolution of the flux ratio between the soft and hard spectral components in the XRP 1A~1118$-$61. To investigate this dependence in \jsevent, here we applied {\tt cflux} to the corresponding {\tt comptt} components to evaluate their X-ray fluxes. The flux ratio $F_{\rm low}/ F_{\rm high}$ shows a clear variation across luminosity states, with values of $0.41\pm0.02$, $0.31\pm0.01$ and $0.45\pm0.01$ in the low, middle, and high states, respectively (see Table~\ref{tab:j17544_flux_resolved_spec}). We applied the model proposed by \cite{Zalot2026} to fit these spectra and the fits yield intersection energies of $E_{\rm int}=7.4\pm0.2$~keV and $7.1\pm0.3$~keV for the middle and low states, respectively. We did not estimate the intersection energy of the high state since it has a high luminosity ($>10^{36}$~\lum), outside the range considered by \cite{Zalot2026}. Using the detected cyclotron line energy of $E_{\rm cyc}=26.6^{+1.2}_{-0.6}$~keV and the intersection energy averaged over these two states, $E_{\rm int}=7.3\pm0.3$~keV, the ratio between $E_{\rm cyc}$ and $E_{\rm int}$ is $3.6\pm0.2$. We note that $E_{\rm cyc}$ and $E_{\rm int}$ were measured in different luminosity states. 
This approach is consistent with the methodology used in BeXBs \citep{Zalot2026}, where $E_{\rm int}$ measured in quiescence was compared to $E_{\rm cyc}$ values obtained during bright outbursts (typically at $L \gtrsim 10^{37}~{\rm erg~s}^{-1}$).  In Fig.~\ref{fig:j17544_flux_resoved_spec}, we plot the results of the spectral fitting at different flux states. The best-fit parameters are shown in Table~\ref{tab:j17544_flux_resolved_spec}.

\begin{table*}
\centering
\caption{Best-fit parameters of \jsevent for flaring and flux-resolved spectra.}
\label{tab:j17544_flux_resolved_spec}
\renewcommand{\arraystretch}{1.3}
\begin{tabular}{lcccc}
\hline
\hline
Parameter & Flaring\tablefootmark{a} & Low state & Middle state & High state \\
\hline
$C_{\rm FPMB}$ & $0.97\pm0.01$ & -- & $0.97\pm0.01$ & $0.98\pm0.01$ \\
$C_{\rm PN}$ & $1.13\pm0.01$ & $1.15_{-0.005}^{+0.014}$ & $0.93\pm0.02$ & $0.90\pm0.01$ \\
$C_{\rm MOS1}$ & $0.71\pm0.01$ & $1.04\pm0.003$ & $0.90\pm0.02$ & $0.83\pm0.03$ \\
$N_{\rm H}$ ($10^{22}$ cm$^{-2}$) & $1.5_{-0.1}^{+0.2}$ & $0.69_{-0.02}^{+0.01}$ & $1.4_{-0.2}^{+0.1}$ & $1.3_{-0.3}^{+0.2}$ \\
$T_{0}$ (keV) & $0.33_{-0.07}^{+0.05}$ & $1.01_{-0.07}^{+0.01}$ & $0.25\pm0.09$ & $0.4\pm0.1$ \\
$kT_{\rm e, low}$ (keV) & $1.50_{-0.05}^{+0.09}$ & $1.54_{-0.07}^{+0.01}$ & $1.25_{-0.04}^{+0.04}$ & $1.59_{-0.08}^{+0.10}$ \\
$\tau_{\rm low}$ & $36_{-3}^{+4}$ & $6.3_{-0.2}^{+4.7}$ & $56_{-7}^{+10}$ & $35_{-5}^{+4}$ \\
Norm$_{\rm low}$ ($10^{-3}$) & $12.9\pm0.1$ & $0.309\pm0.003$ & $5.6_{-0.5}^{+0.3}$ & $64_{-4}^{+5}$ \\
$kT_{\rm e, hard}$ (keV) & $9.1_{-0.9}^{+1.3}$ & $6.9_{-0.1}^{+0.9}$ & $15_{-2}^{+4}$ & $7.6_{-0.9}^{+1.1}$ \\
$\tau_{\rm hard}$ & $100.0$ (fixed) & $100.0$ (fixed) & $100.0$ (fixed) & $100.0$ (fixed) \\
Norm$_{\rm hard}$ ($10^{-3}$) & $3.4_{-0.3}^{+0.4}$ & $0.030\pm0.001$ & $2.27_{-0.04}^{+0.07}$ & $16.3_{-0.6}^{+1.1}$ \\

$E_{\rm gau}$ (keV) & $6.38_{-0.07}^{+0.05}$ & -- & $6.37_{-0.06}^{+0.05}$ & $6.41_{-0.08}^{+0.06}$ \\
$\sigma_{\rm gau}$ (keV) & $0.1$ (fixed) & -- & $0.1$ (fixed) & $0.1$ (fixed) \\
$Norm_{\rm gau}$ ($10^{-4}$) & $0.95_{-0.03}^{+0.02}$ & -- & $0.5\pm0.2$ & $6\pm2$ \\

$E_{\rm cyc, 1}$ (keV) & $26.6_{-0.6}^{+1.2}$ & $26.6$ (fixed) & $26.6$ (fixed) & $26.6$ (fixed) \\
$\sigma_{\rm cyc, 1}$ (keV) & $6.4_{-0.4}^{+0.7}$ & $6.4$ (fixed) & $6.4$ (fixed) & $6.4$ (fixed) \\
$d_{\rm cyc, 1}$ & $1.6_{-0.2}^{+0.3}$ & $1.7_{-0.2}^{+0.4}$ & $2.0\pm0.2$ & $1.4\pm0.2$ \\
$E_{\rm cyc, 2}$ (keV) & $53.2$ (fixed) & $53.2$ (fixed) & $53.2$ (fixed) & $53.2$ (fixed) \\
$\sigma_{\rm cyc, 2}$ (keV) & $12.5_{-1.0}^{+0.7}$ & $12.5$ (fixed) & $12.5$ (fixed) & $12.5$ (fixed) \\
$d_{\rm cyc, 2}$ & $7\pm1$ & $<0.02$ & $7.6_{-0.9}^{+0.7}$ & $6.4_{-0.9}^{+1.0}$ \\

$F_{\rm low}/F_{\rm hard}$\tablefootmark{b} & -- & ($0.14\pm0.01$)/($0.34\pm0.01$) & ($3.1\pm0.1$)/($10.1\pm0.2$) & ($50\pm1$)/($112\pm2$) \\
$L_{\rm 0.5-60~keV}$ ($10^{34}$~\lum) & $21.0\pm0.2$ & $0.34\pm0.01$ & $9.1\pm0.2$ & $112\pm1$ \\
\hline
$\chi^2$ (d.o.f) & $1653.9$ ($1604$) & $1233.4$ ($1168$) & $1099.0$ ($1006$) & $1174.9$ ($1146$) \\
\hline
\end{tabular}
\tablefoot{The {\it NuSTAR}/FPMA constant is fixed at 1. The luminosities are unabsorbed values, calculated by assuming an isotropic source at a distance of 2.4~kpc. \\
\tablefoottext{a}{The time interval used to extract the spectra during the flaring period is marked by the green shaded region in Fig.~\ref{fig:lc_j17544}.}
\tablefoottext{b}{The unit of flux is $10^{-11}$\flux, estimated for the 0.5--60~keV energy range.} }
\end{table*}


\begin{figure}
\centering
\includegraphics[width=0.95\linewidth]{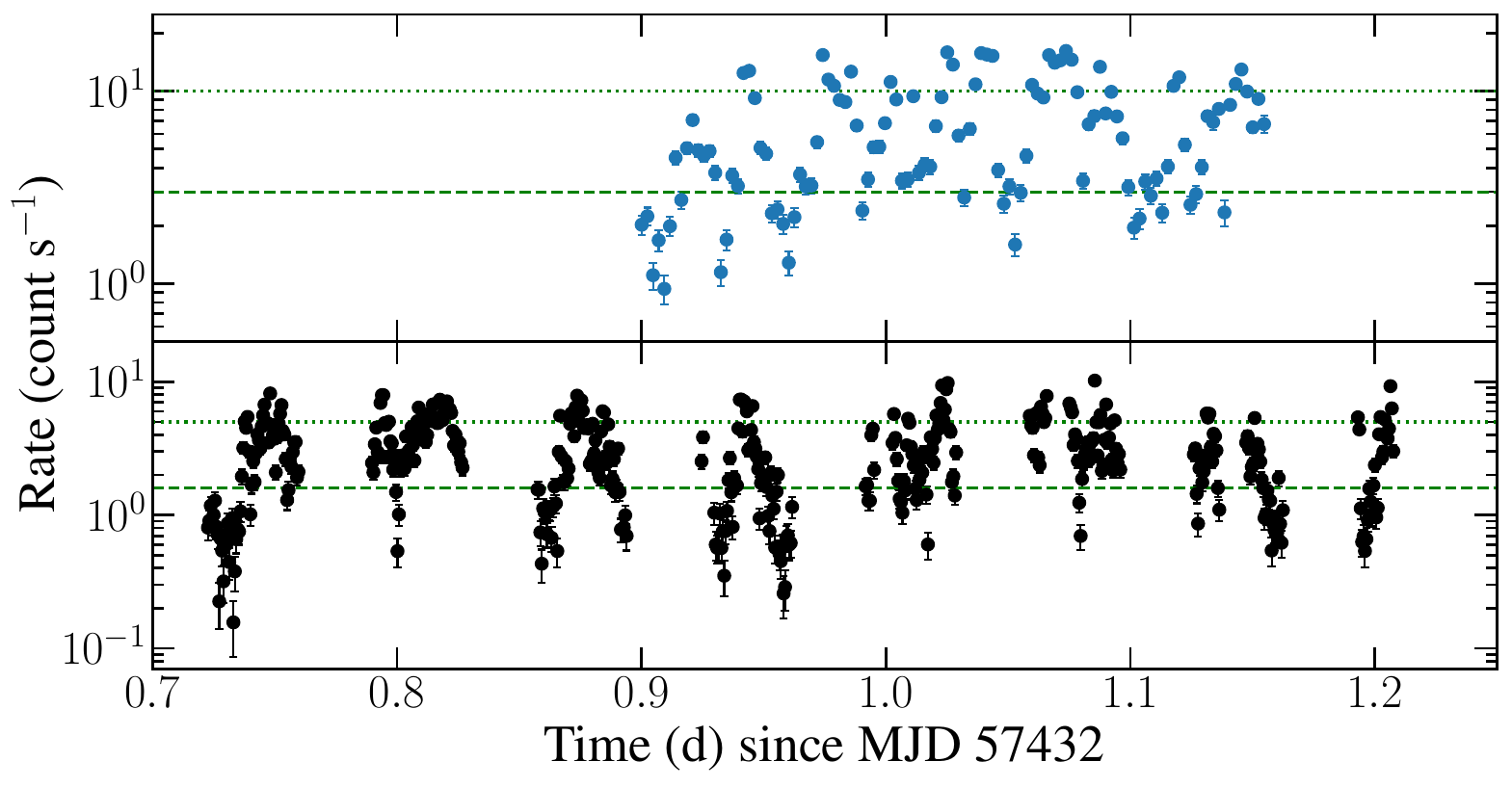} 

\caption{Background-subtracted light curves of \jelev observed by {\it XMM-Newton}/EPIC-PN (top; 0.3--10~keV) and {\it NuSTAR}/FPMA (bottom; 3--79~keV). The green dashed and dotted lines indicate the count-rate thresholds (3.0 and 9.8~count~s$^{-1}$ for {\it XMM-Newton}; 1.6 and 5.0~count~s$^{-1}$ for {\it NuSTAR}), defining the low and high flux states used for the flux-resolved spectroscopy.}
\label{fig:lc_j11215}
\end{figure}

\begin{figure}
\centering
\includegraphics[width=0.95\linewidth]{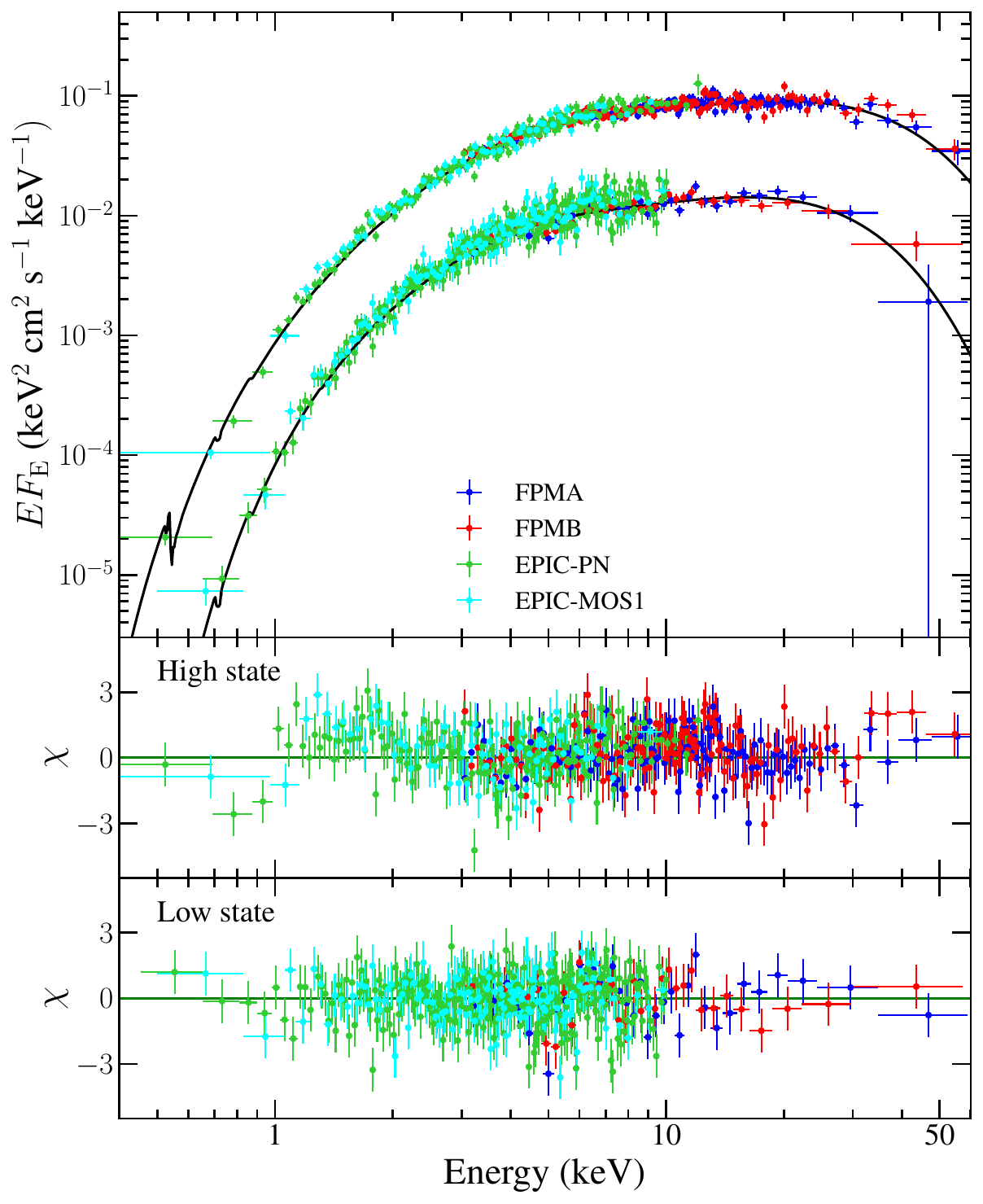}
\caption{Unfolded spectra of \jelev, extracted from different state with luminosities of $1.7\times10^{36}$~\lum (high state) and $1.2\times10^{35}$~\lum (low state), respectively, estimated by using a {\tt const$\times$tbabs$\times$comptt} model. Corresponding residuals are shown in the lower panel.}
\label{fig:j11215_flux_resolved_spec}
\end{figure}

\subsection{\jelev}

For flux-averaged spectra of \jelev, the joint {\it XMM-Newton} and {\it NuSTAR} spectra can be well described by a single phenomenological cutoff power-law model ({\tt cutoffpl}) or a thermal Comptonization model like {\tt comptt} modified by a photoelectric absorption component {\tt tbabs}, which is consistent with results obtained by \cite{Sidoli2017}. The authors reported negative residuals in the 15--40~keV energy range, yet an additional {\tt cyclabs} component did not provide sufficient statistical significance to confirm the presence of a cyclotron line. Since the single continuum model provides a good fit to the average spectrum, we did not perform any further searches for cyclotron lines. 
To verify the conclusion that the spectrum of  this SFXT evolves into a two-hump morphology in the low state, we extracted spectra from different intensity levels, defined by {\it NuSTAR} count rates below 1.6~count~s$^{-1}$ and above 5~count~s$^{-1}$, corresponding to {\it XMM-Newton} count rates below 3.0~count~s$^{-1}$ and above 9.8~count~s$^{-1}$, respectively (see Fig.~\ref{fig:lc_j11215}). However, unlike the cases of \jsixt and \jsevent which exhibit a distinct two-hump morphology in the low state, the spectrum of \jelev shows a smooth broadband continuum in both the low and high states, without the need for  additional soft or hard components. We tried to describe the spectra using a {\tt comptt+comptt} model, or a {\tt bbodyrad+cutoffpl} model. However, neither of these models significantly improves the fit, and no absorption features are found in the residuals between 0.5 and 60~keV. We therefore fitted the spectra using only a {\tt comptt} model. The fitted spectra and the residuals are displayed in Fig.~\ref{fig:j11215_flux_resolved_spec}, and the best-fit parameters are summarized in Table~\ref{tab:j11215_flux_resolved_spec}.

\begin{table}
\centering
\caption{Best-fit spectral parameters of \jelev for low and high states}
\label{tab:j11215_flux_resolved_spec}
\renewcommand{\arraystretch}{1.3}
\begin{tabular}{lcc}
\hline
\hline
Parameter & Low State & High State \\
\hline
$C_{\rm FPMB}$ & $1.03\pm0.05$ & $0.97\pm0.02$ \\
$C_{\rm PN}$ & $0.77_{-0.03}^{+0.04}$ & $0.97_{-0.02}^{+0.03}$ \\
$C_{\rm MOS1}$ & $0.82_{-0.03}^{+0.05}$ & $0.97_{-0.03}^{+0.04}$ \\
$N_{\rm H}$ ($10^{22}$ cm$^{-2}$) & $0.8\pm0.1$ & $0.34\pm0.04$ \\
$T_{0}$ (keV) & $0.97_{-0.05}^{+0.04}$ & $1.11_{-0.02}^{+0.03}$ \\
$kT_{\rm e}$ (keV) & $5.9_{-0.6}^{+0.9}$ & $8.1_{-0.4}^{+0.6}$ \\
$\tau$ & $11\pm1$ & $8.4_{-0.5}^{+0.3}$ \\
Norm ($10^{-3}$) & $0.8\pm0.1$ & $3.7_{-0.3}^{+0.2}$ \\
$L_{\rm 0.5-60~keV}$ ($10^{34}$~\lum)  & $12.4_{-0.4}^{+0.5}$ & $168.5_{-3.3}^{+3.4}$  \\
\hline
$\chi^2$ (d.o.f) & $1476.0$ ($1711$) & $1530.8$ ($1425$) \\
\hline
\end{tabular}
\tablefoot{The {\it NuSTAR}/FPMA constant is fixed at 1. The luminosities are unabsorbed values, calculated by assuming an isotropic source at a distance of 6.5~kpc.}
\end{table}

\section{Discussion} 
\label{sec:dis}

SFXTs are a distinct subclass of HMXBs characterized by their persistent low luminosity and rare, intense flares. This behavior of the systems make it an ideal targets for studying the physics of accreting magnetized NSs in the low-luminosity state. In this study, we performed a broadband spectral analysis of the SFXTs \jsixt, \jsevent, and \jelev using archival {\it XMM-Newton}, NICER, and {\it NuSTAR} observations.
\subsection{Magnetic field strength of the NS}
%
SFXTs are believed to host a highly magnetized NS. Cyclotron lines provide a direct measurement of the magnetic field strength in these systems.
Our spectral fitting of \jsixt supports the presence of a CRSF at $\sim$50~keV. The best-fit line energy of $E_{\rm cyc}=53.9_{-1.8}^{+2.3}$~keV implies a NS magnetic field strength of $B\approx(6.0^{+0.3}_{-0.2})\times10^{12}$~G, obtained using the formula $E_{\rm cyc} \simeq \, 11.6 \times B_{12}/(1+z)$\,keV, where $E_{\rm cyc}$ is the centroid line energy, 
$B_{12}$ the magnetic field strength in units of $10^{12}$\,G and $z=0.3$ is the gravitational redshift for a canonical NS ($R_{\rm NS}=10$~km and $M_{\rm NS}=1.4M_{\odot}$). In contrast, using both {\tt cutoffpl} and {\tt comptt} models, we find no significant absorption structure in the flux-averaged or the flux-resolved spectra of \jelev. This is consistent with the non-detection reported by \citet{Sidoli2020}.

The magnetic field of \jsevent has been estimated to range from $(1.45\pm0.03)\times10^{12}$~G \citep{Bhalerao2015} to $\gtrsim10^{14}$~G \citep{Bozzo2008} based on a cyclotron line at 17.0$\pm0.3$~keV and the magnetic gating mechanism used to explain its flare behavior, respectively. Using a double-Comptonization continuum, we detected an absorption line at $26.6^{+1.2}_{-0.6}$~keV, which is higher than the value reported by \cite{Bhalerao2015}. Although the two spectra were fitted using different continuum models, the substantial shift in line energy is unlikely to be explained solely by model dependence. If we assume that both lines are real, the line energy increased by $\Delta E \approx10$~keV within two years. Long-term changes in line energies have been observed in several accreting XRPs, e.g., \mbox{Her~X-1}, \mbox{Vela~X-1}, and \mbox{Cen~X-3} \citep[e.g.,][]{Ji2019}, where the line centroid has evolved over years to decades. Such variations are generally interpreted as a change in the local magnetic configuration \citep{Staubert2014}. Two non-harmonically related cyclotron lines have been observed simultaneously in several sgXBs (e.g., GX~301$-$2, \citealt{Furst2018}; 4U~1700$-$37, \citealt{Xiao2024}), which were explained by changes in the geometry of the line-forming region. On the other hand, we note that the 2013 observation of \jsevent was performed at an X-ray luminosity of $\sim2.4\times10^{34}$~\lum\ ($F_{3-50\rm~keV}=3.5 \times10^{-11}$~\flux, assuming a distance of 2.4 kpc), whereas our observations show a luminosity approximately an order of magnitude higher ($2.1\times10^{35}$~\lum). 
A higher luminosity can provide stronger radiation-pressure support and reduce the velocity of the accreting plasma before it reaches the NS surface.
The resonant scattering of X-ray photons by the moving plasma in the accretion channel is affected by the Doppler effect, so that a slower inflow can shift the observed cyclotron line towards higher energies.
This mechanism was originally proposed by
\citep{Mushtukov2015b} and has recently been confirmed by numerical simulations of radiative transfer in the accretion channel \citep{2026A&A...709A.247F,2026MNRAS.551g1517M}.
A short-timescale increase of the line energy was also reported for 1A~0535+262 during its 2020 giant outburst \citep{Kong2021}, where $E_{\rm cyc}$ varied approximately linearly at a rate of $\Delta E_{\rm cyc} = 0.071$~keV~d$^{-1}$, and was interpreted in terms of changes in the structure of inner accretion disk \citep{Doroshenko2017}. For \jsevent, the flare around MJD 57102.66 may have modified the accretion flow and triggered a change in the observed cyclotron line. This requires further observations to verify and is therefore beyond the scope of the present work.

%

%



\subsection{Two-hump spectrum in SFXTs}

Using a double-Comptonization model, we reported the identification of two-hump spectra in the low-luminosity states of two SFXTs \jsixt (Fig.~\ref{fig:j16418_flux_resoved_spec}) and \jsevent  (Fig.~\ref{fig:j17544_flux_resoved_spec}), consistent with the recent detection in the 
SFXT IGR~J00370+6122 \citep{Ducci2026}. This spectral feature has been widely observed in BeXBs, including both transient sources and persistent sources (e.g., 1A~0535+262, \citealt{Tsygankov2019a}; GX~304$-$1, \citealt{Tsygankov2019b}; GRO~J1008$-$57, \citealt{Lutovinov2021}; MAXI~J0655$-$013, \citealt{Malacaria2026}; 1A~1118$-$61, \citealt{Salganik2026}; X~Persei, \citealt{Doroshenko2012}; 2RXP~J130159.6$-$635806, \citealt{Salganik2025};  IGR~J06074+2205, \citealt{Salganik2026a}, see also \citealt{Zalot2026} for more sources). 

The spectral fitting results show that the luminosities of all SFXTs in our sample were $\lesssim10^{36}$~\lum\ during the observation periods, indicating that the sources are expected to be in the subsonic (settling) accretion regime and a hot quasi-static shell should form around the magnetosphere \citep{Shakura2012}. The plasma enters the magnetosphere of the NS through Rayleigh-Taylor instabilities after cooling. In classical BeXBs, by contrast, the NS is fed through an accretion disc. The disc is truncated at the magnetospheric radius, where the material is channelled by the magnetic field lines onto the polar caps \citep[for a review, see][]{MushtukovTsygankov2024}. Our results suggest the observed two-hump morphology is a universal feature of accreting magnetized NSs at low luminosity state, independent of accretion mode.

Theoretical models of low-luminosity accreting XRPs \citep{Mushtukov2021,Sokolova-Lapa2021} explain the two-hump spectral structure in terms of radiation emerging from an accretion-heated, strongly magnetized NS atmosphere.
Accretion provides energy for both thermal radiation and cyclotron photon production, while resonant scattering plays an important role in shaping the emerging spectrum, particularly its high-energy component.
These two theoretical models adopt different physical approximations and numerical methods for treating radiative transfer, leading to differences in the relative contributions of individual radiation processes.
In particular, \citet{Sokolova-Lapa2021} demonstrated that the observed spectral shape can be reproduced with a relatively smaller contribution from cyclotron photons generated by collisional excitation.
These differences do not imply fundamentally distinct physical scenarios, and the relative importance of the various processes remains dependent on the adopted approximations and atmospheric conditions.

The best-fit parameters of the double-Comptonization spectral model support to the heated-atmosphere scenario proposed for the low-luminosity regime. The temperature of the seed photons, $kT_0 \simeq$0.2--1.4~keV (see Table~\ref{tab:j16418_spec} and \ref{tab:j17544_flux_resolved_spec}), is consistent with the expected blackbody temperature of hotspots around the magnetic pole regions. The electron temperature of the low-energy component, $kT_{\rm e,low}\simeq$1.2--5.6~keV, matches the values found in other low-luminosity pulsars, implying that the low-energy hump arises from thermal Comptonization in the heated atmosphere. We note that the electron temperature of the hard component ($kT_{\rm e,hard}$) is strongly degenerate with the optical depth, which we fixed at $\tau_{\rm hard}=100$. They are therefore used only for a phenomenological description of the spectrum, without inferring the underlying physical scenario.
%

\begin{figure}
\centering
\includegraphics[width=0.9\linewidth]{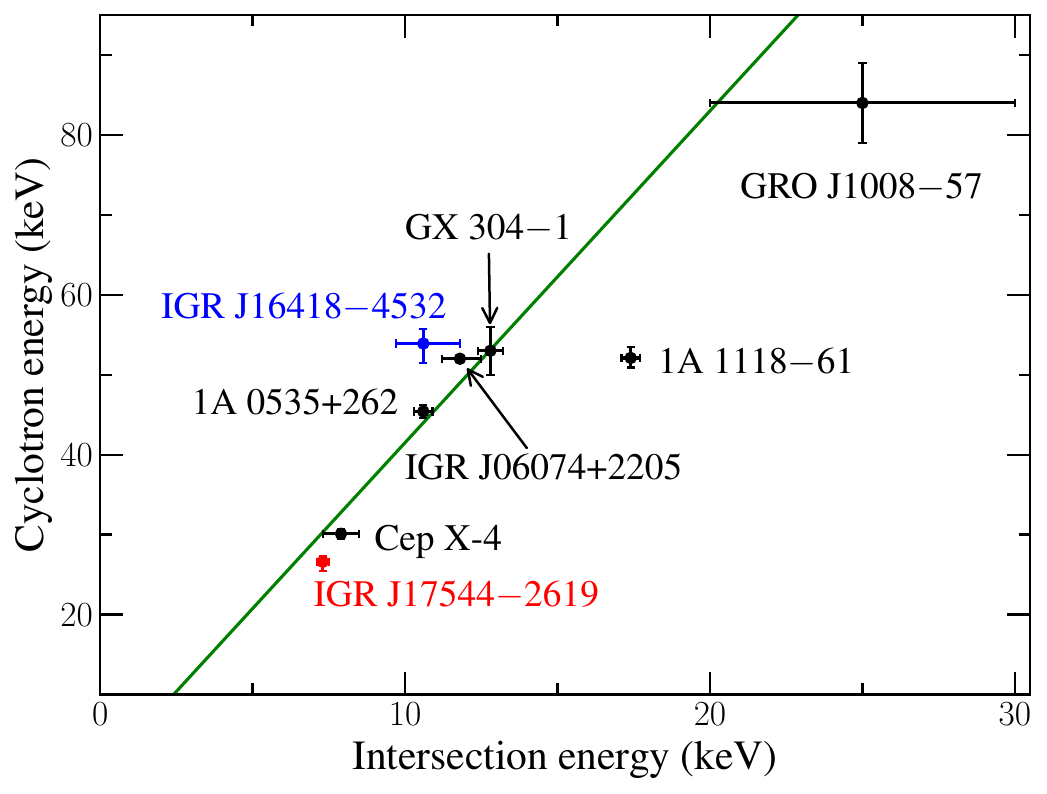} 

\caption{Relation between cyclotron line energy ($E_{\rm cyc}$) and intersection energy ($E_{\rm int}$) for a set of XRPs. 
Data points for \jsixt  and \jsevent obtained in this paper are shown in blue and red, respectively. Other points (in black) for BeXBs are 
from \citet{Zalot2026}, \citet{Salganik2026a}, and \citet{Salganik2026}. The green line represents a linear function of $E_{\rm cyc}$ and $E_{\rm int}$ with a slope of 4.15.
\label{fig:E_int_vs_E_cyc}}
\end{figure}

%
\jsixt shows the two-hump shape only in its low state ($9.9\times10^{34}$~\lum), evolving to a single-component continuum at $1.5\times 10^{36}$~\lum. \jsevent, in contrast, retains the two-hump morphology over nearly three orders of magnitude in luminosity ($10^{33}-10^{36}$~\lum), with the flux ratio $F_{\rm low}/F_{\rm high}$ ranging from $0.31\pm0.01$ to $0.45\pm0.01$ across different flux states. This evolution may be consistent with that observed in 1A~1118$-$61, which was attributed to changes in the efficiency of exciting electrons to higher Landau levels at different luminosities and magnetic fields \citep{Salganik2026}.
\cite{Zalot2026} reported that the intersection energy exhibits a linear relationship with cyclotron line energy ($E_{\rm cyc} = (4.15\pm0.11)\times E_{\rm int}$) in the BeXB sample. We note that the individual ratios reported by \citet{Zalot2026} range from $3.3\pm0.6$ for GRO~J1008$-$57 to $4.3\pm0.2$ for 1A~0535+26. For \jsixt and \jsevent, we obtained $E_{\rm cyc}/E_{\rm int}$ ratios of $5.1_{-0.6}^{+0.5}$ and $3.6_{-0.1}^{+0.2}$, respectively. Therefore, despite some scatter in the individual ratios, our results suggest that the linear correlation appears to remain applicable to SFXTs (Fig.~\ref{fig:E_int_vs_E_cyc}). This may also naturally explain the absence of a two-hump spectral shape in \jelev, which has been suggested to host a NS with a magnetic field of $2\times 10^{12}$~G \citep{Sidoli2017}. At such a low field strength, resonant redistribution is expected to flatten the spectrum \citep{Gorban2022}, causing the two humps to merge. However, we note that the discussion of the two-hump morphology in SFXTs is currently based on only one observation per source, and the accurate physical features of the high-energy component are still lacking. Future observations, particularly those with higher sensitivity and broader X-ray energy coverage, will help us to better understand the properties of SFXTs in low-luminosity states.

\section{Conclusions}

We performed a broadband spectral analysis of three SFXTs, \jsixt, \jsevent, and \jelev, using archival {\it XMM-Newton}, NICER, and {\it NuSTAR} observations covering the $\sim$0.5--79~keV energy range. Our main findings are summarized as follows.

\begin{itemize}

\item We confirmed the presence of a CRSF at $E_{\rm cyc}=53.9^{+2.3}_{-1.8}$~keV in \jsixt, implying a NS magnetic field strength of $\sim$$6\times10^{12}$~G. For \jsevent, we detected a CRSF at $E_{\rm cyc}=26.6^{+1.2}_{-0.6}$~keV, significantly higher than the previously reported value of $\sim$17~keV. This shift may be related to changes in the accretion flow and the geometry of the line-forming region at different luminosities. In \jelev, we did not find any significant evidence of cyclotron absorption features in either the flux-averaged or flux-resolved spectra.

\item We identified the characteristic two-hump spectral shape in \jsixt and \jsevent at luminosities $\lesssim10^{35}$~\lum. In \jsixt, the two-hump morphology is observed only in the low-luminosity state, whereas \jsevent exhibits two distinct spectral components over a much broader luminosity range. These results suggest that the two-hump spectral morphology observed in BeXBs may also be a common feature of magnetized NSs in wind-fed systems, despite differences in their accretion mechanisms.

\item For \jsixt and \jsevent, we obtained $E_{\rm cyc}/E_{\rm int}$ ratios of $\sim5.1$ and $\sim3.6$, respectively, broadly consistent with the empirical relation established for BeXBs. This suggests that the relation may also hold for SFXTs, providing a potential method for estimating the NS magnetic field strength when cyclotron lines are not directly detectable. In contrast, \jelev shows no evidence of a two-hump spectral shape, possibly indicating a weaker magnetic field, although its strength remains poorly constrained.

\end{itemize}

Our findings suggest that the two-hump spectral morphology is a common feature of accreting magnetized NSs at low luminosities, regardless of the accretion mode. Further broadband observations of SFXTs can test the applicability of the $E_{\rm int}$--$E_{\rm cyc}$ relation to a larger sample of sources.

\begin{acknowledgements}
This research was supported by the  Research Council of Finland Centre of Excellence in Neutron-Star Physics (grant 374064). HX thanks the support from the China Scholarship Council (CSC: 202406380048). This research has made use of software provided by the High Energy Astrophysics Science Archive Research Centre (HEASARC), which is a service of the Astrophysics Science Division at NASA/GSFC and the High Energy Astrophysics Division of the Smithsonian Astrophysical Observatory. This work is based on observations obtained with {\it NuSTAR}, NICER, and {\it XMM-Newton}.
\end{acknowledgements}

\bibliographystyle{aa}
\bibliography{allbib}

\end{document}